\documentclass[
    ,final            % use final for the camera ready runs
  ]
  {aipproc}

\layoutstyle{6x9}

\usepackage{amsmath}

\newcommand{\lb}{\left(}
\newcommand{\rb}{\right)}

\begin{document}

\title{Local covariant density functional constrained by the relativistic Hartree-Fock theory}


\author{H. Z. Liang}{
  address={State Key Lab Nucl. Phys. {\rm \&} Tech., School of Physics, Peking University, Beijing 100871, China},
  altaddress={RIKEN Nishina Center, Wako 351-0198, Japan}
}

\author{J. Meng}{
  address={State Key Lab Nucl. Phys. {\rm \&} Tech., School of Physics, Peking University, Beijing 100871, China},
  altaddress={School of Physics and Nuclear Energy Engineering, Beihang University, Beijing 100191, China}
}

\author{P. Ring}{
  address={Physik Department, Technische Universit\"at M\"unchen, D-85747 Garching, Germany},
  altaddress={State Key Lab Nucl. Phys. {\rm \&} Tech., School of Physics, Peking University, Beijing 100871, China}
}

\author{X. Roca-Maza}{
  address={INFN, Sezione di Milano, via Celoria 16, I-20133 Milano, Italy}
}

\author{P. W. Zhao}{
  address={State Key Lab Nucl. Phys. {\rm \&} Tech., School of Physics, Peking University, Beijing 100871, China}
}

\begin{abstract}
The recent progress in the localized covariant density functional constrained by the relativistic Hartree-Fock theory is briefly presented by taking the Gamow-Teller resonance in $^{90}$Zr as an example.
It is shown that the constraints introduced by the Fock terms into the particle-hole residual interactions are straight forward and robust.
\end{abstract}

\maketitle

%\section{Introduction}

\textbf{Introduction} During the past decades, the covariant density functional theory (CDFT) has received wide attention due to its successful descriptions of both ground-state and excited-state properties of nuclei all over the nuclear chart.
In this report, we will mainly focus on our recent progress in the localized covariant density functional constrained by the relativistic Hartree-Fock (RHF) theory \cite{Liang2012b}.

Research on quantum mechanical many-body problems is essential in many fields of modern physics.
By reducing the many-body problem formulated in terms of $N$-body wave functions to the one-body level with local density distributions $\rho(\mathbf{r})$, the density functional theory (DFT) of Kohn and Sham \cite{Kohn1965} has accomplished great success.
In particular, this theory states that there always exists a local single-particle potential $V_{\rm KS}(\mathbf{r})$ so that the exact ground-state density of the interacting system can be reproduced by non-interacting particles moving in such local potential.

The DFT has been also widely used in nuclear physics since the 1970s \cite{Bender2003}.
In particular, the Lorentz symmetry taken into account in the CDFT puts stringent restrictions on the number of parameters in the corresponding functionals without reducing the quality of the agreement with experimental data.
During the past decades, the CDFT has been successful in describing a large variety of nuclear phenomena, including the equation of state in nuclear matter, ground-state properties of finite nuclei, collective rotational and vibrational excitations, as well as fission landscapes and low-lying spectra of transitional nuclei involving quantum phase transitions.
The readers are referred to Refs.~\cite{Ring1996,Vretenar2005,Meng2006,Paar2007,Niksic2011} for recent reviews.

Part of these successes could only be achieved because the underlying functionals depend only on local densities, i.e., the relativistic Hartree (RH) scheme.
However, a few common problems have been found in the isovector channels of the widely used RH parametrizations.
Since a few years, the RHF calculations are possible for spherical nuclei all over the nuclear chart \cite{Long2006}.
In contrast to the simple Hartree calculations, they reproduce successfully the proton-neutron effective mass splitting \cite{Long2006} (at the Hartree level it can be reproduced only if a isovector-scalar meson is included \cite{Roca-Maza2011}), shell structure evolutions \cite{Long2007,Long2008,Tarpanov2008,Long2009,Moreno-Torres2010}, and spin-isospin resonances \cite{Liang2008,Liang2009,Liang2012}.
However, since the RHF theory introduces non-local potentials $V_{\rm HF}(\mathbf{r}, \mathbf{r}')$, its theoretical framework is much more involved and the simplicity of the Kohn-Sham scheme is lost.
This restricts the applicability of deformed RHF theory to a few very light nuclei \cite{Ebran2011} and it prevents the treatment of effects beyond mean-field.
Therefore, it is highly desirable to find a covariant density functional based on only local potentials, yet keeping the merits of the exchange terms.

Recently, we started from the important observation that one of the successful and widely used RHF parametrizations PKO2 \cite{Long2008} includes only three relatively heavy mesons, $\sigma$, $\omega$, and $\rho$, but no pions.
Since the masses of these mesons are relatively heavy, the zero-range reduction for the corresponding finite-range Yukawa propagators becomes a reasonable approximation.
As a step further, with the help of Fierz transformation \cite{Okun1982,Sulaksono2003}, the Fock terms can be expressed as local Hartree terms.
In such way, a Hartree-Fock equivalent covariant density functional can be derived while the constraints introduced by
the Fock terms of the RHF scheme can be taken into account.
This RHF equivalent parametrization thus obtained is called PKO2-H \cite{Liang2012b}.
For details, the formalism for the effective Lagrangian density as well as the zero-range reduction and Fierz transformation can be found in Ref.~\cite{Liang2012b}.

In the following, we will take the Gamow-Teller resonance (GTR) in $^{90}$Zr as an example to illustrate the ideas of the localized RHF equivalent RPA and the main conclusions.

%\section{Localized RHF equivalent RPA}

\begin{figure}
  \includegraphics[width=.50\textwidth]{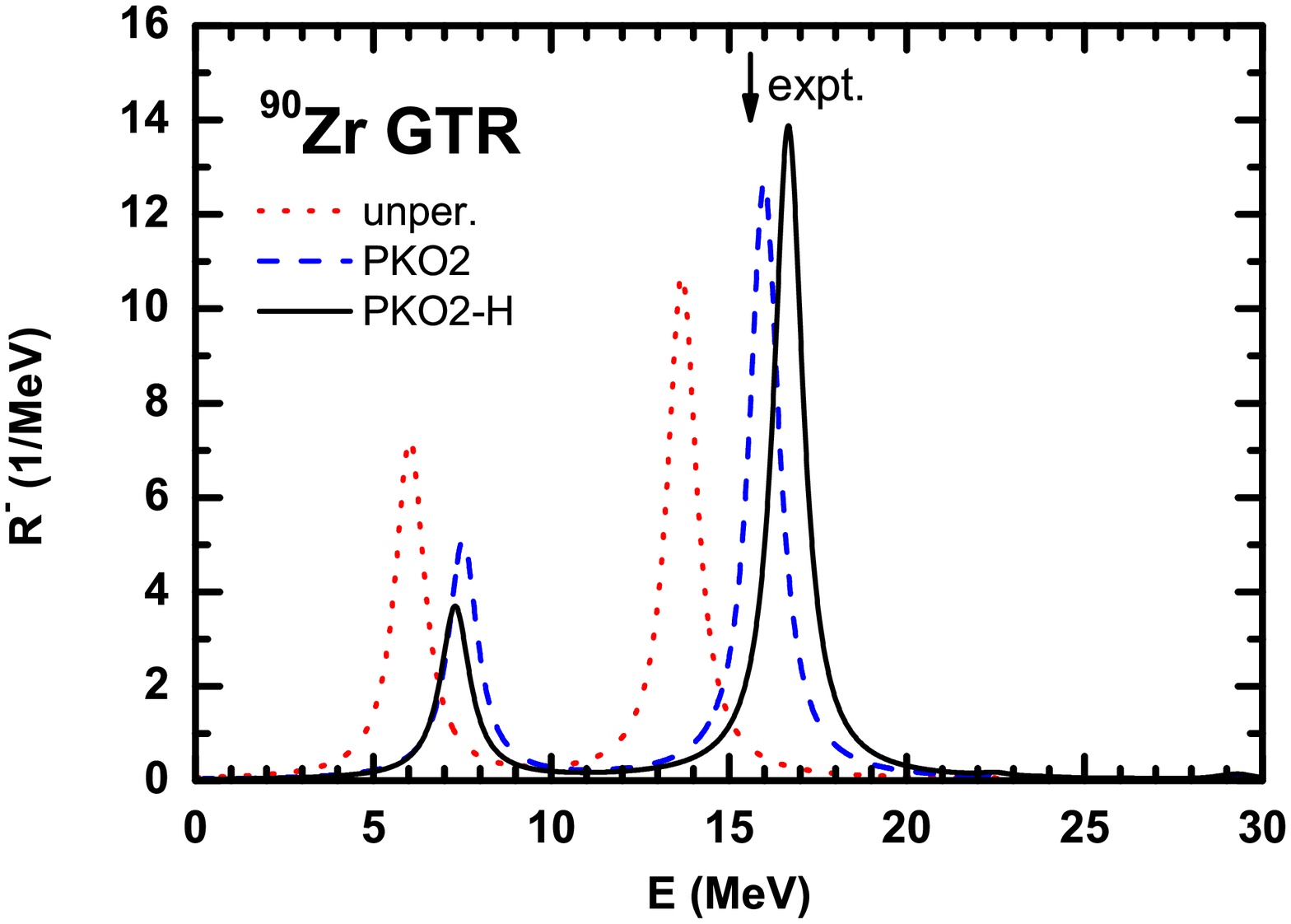}
  \includegraphics[width=.50\textwidth]{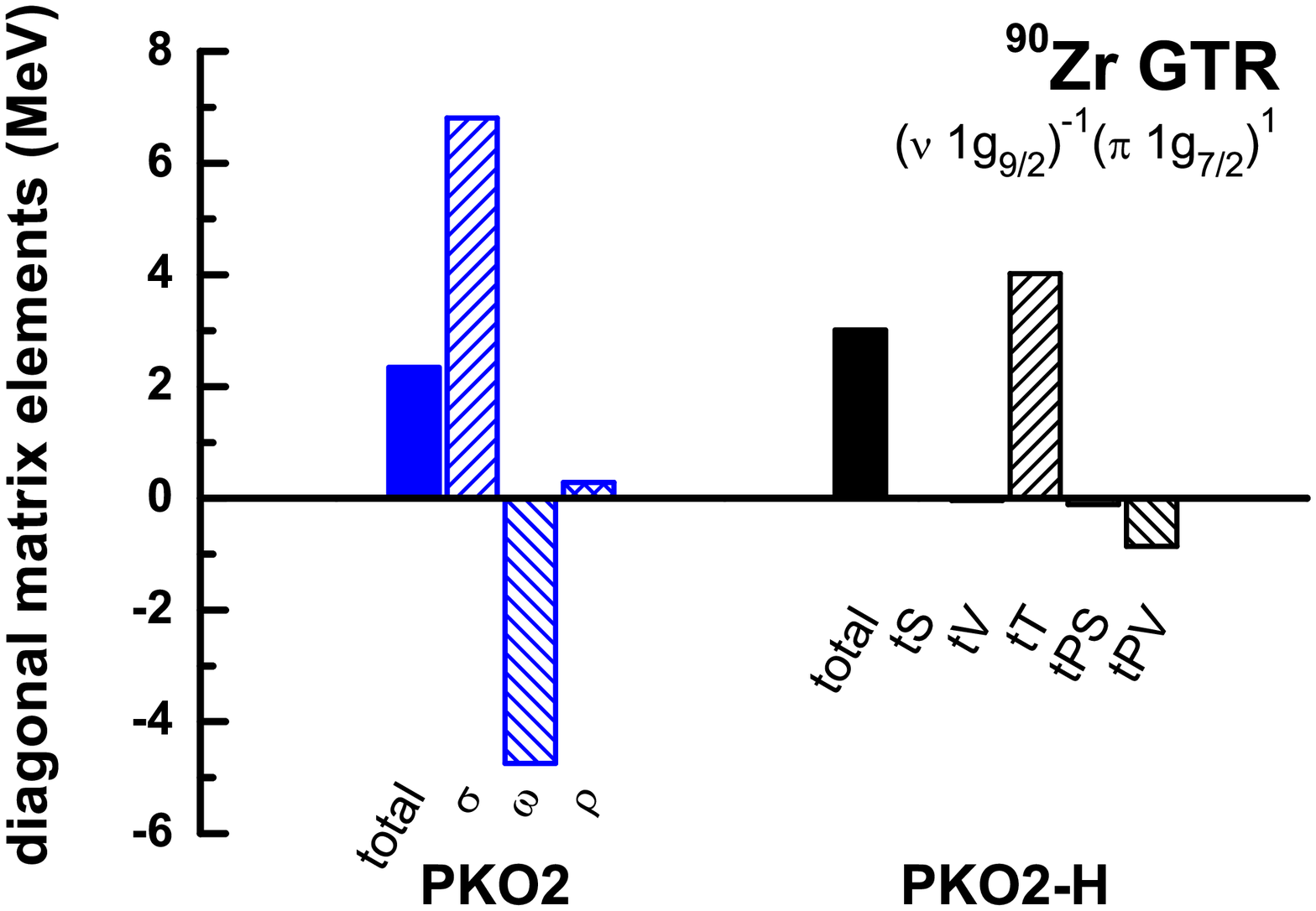}
  \caption{(Color online) Left panel: Strength distribution of GTR in $^{90}$Zr calculated with the \textit{ph} residual interactions of PKO2-H \cite{Liang2012b} with Lorentzian smearing parameter 1 MeV. The unperturbed excitations and the original RHF+RPA \cite{Liang2008} results with PKO2 \cite{Long2008} are shown as the dotted and dashed lines for comparison. The experimental peak energy \cite{Wakasa1997} is denoted by the arrow.
  Right panel: Diagonal matrix elements of \textit{ph} residual interactions for the $(\nu1g_{9/2})^{-1}(\pi1g_{7/2})^1$ configuration. The total matrix elements are decomposed into the contributions from different channels.
    \label{Fig1}}
\end{figure}

\textbf{Results and discussion} In the left panel of Fig.~\ref{Fig1}, the strength distribution of GTR in $^{90}$Zr calculated by the particle-hole (\textit{ph}) residual interactions of PKO2-H is shown as solid line.
It is found that, based on the same unperturbed single-particle spectra calculated by the RHF theory (dotted line), the present results by the localized RHF equivalent RPA calculations are quite similar to those by the original RHF+RPA \cite{Liang2008} calculations (dashed line). The slight difference is due to the zero-range approximation.
The experimental peak energy~\cite{Wakasa1997} is well reproduced.

In order to understand the physical mechanisms, we present in the right panel of Fig.~\ref{Fig1} the diagonal matrix elements of the most important GT \textit{ph} configuration $(\nu 1g_{9/2})^{-1}(\pi 1g_{7/2})^1$.
The total matrix elements (full) are decomposed into the contributions from different channels.

In the original RHF+RPA calculations with PKO2, the \textit{ph} residual interactions are provided by the $\sigma$, $\omega$, and $\rho$ mesons.
Both the direct and exchange terms of the \textit{ph} residual interactions must be taken into account.
It has been demonstrated that the total \textit{ph} strengths are essentially determined by the delicate balance between the $\sigma$ and $\omega$ mesons \cite{Liang2008,Liang2012}.
Since these two mesons are isoscalar, they can contribute via the Fock terms only, and correspondingly the effects of $\sigma$ and $\omega$ mesons are repulsive and attractive, respectively.

By taking the zero-range approximation of the finite-range meson-exchange Yukawa propagators and using the Fierz transformation, the \textit{ph} interactions provided by the $\sigma$, $\omega$, and $\rho$ mesons can be mapped into ten possible channels, i.e., the scalar ($S$), vector ($V$), tensor ($T$), pseudoscalar ($PS$), and pseudovector ($PV$) channels with the isoscalar and isovector ($t$) nature.
For the charge-exchange modes, only the isovector channels contribute, since just the direct terms of the \textit{ph} residual interactions are taken into account when the Fierz transformation has been preformed.

Specifically, for the charge-exchange spin-flip modes like GTR, the dominant residual interactions are represented by the matrix elements of the spin-isospin operator $[\boldsymbol\sigma\vec\tau]\cdot[\boldsymbol\sigma\vec\tau]$ between the large components of Dirac spinors.
They are composed of $[\sigma^{ij}\vec\tau]\cdot[\sigma_{ij}\vec\tau]$ and $[\gamma_5\gamma^{i}\vec\tau]\cdot[\gamma_5\gamma_{i}\vec\tau]$ terms with the coefficients
\begin{equation}
    2\alpha^{\rm H}_{tT} = \frac{1}{8}\frac{g^2_\sigma}{m^2_\sigma}
    \qquad\mbox{and}\qquad
    -\alpha^{\rm H}_{tPV} = \frac{1}{8}\frac{g^2_\sigma}{m^2_\sigma} - \frac{1}{4}\frac{g^2_\omega}{m^2_\omega} + \frac{1}{4}\frac{g^2_\rho}{m^2_\rho},
\end{equation}
respectively, where $\sigma^{ij},\gamma^{i},\gamma_5$ are Dirac matrices, $g_x$ and $m_x$ ($x=\sigma,\omega,\rho$) are the meson-nucleon coupling strengths and meson masses in the original RHF parametrization PKO2 \cite{Long2008}, while $\alpha^{\rm H}$ are the Hartree equivalents.
Correspondingly, the net contribution is then approximately proportional to
\begin{equation}\label{Eq2}
    2\alpha^{\rm H}_{tT}-\alpha^{\rm H}_{tPV}
    \approx \frac{1}{4}\lb \frac{g^2_\sigma}{m^2_\sigma} - \frac{g^2_\omega}{m^2_\omega}\rb,
\end{equation}
as $g^2_\rho\ll g^2_\omega, m^2_\rho\approx m^2_\omega$.

As shown in the right panel of Fig.~\ref{Fig1}, it is true that, within the present localized RHF equivalent RPA approach, the total \textit{ph} strengths are mainly composed of the positive $tT$ and negative $tPV$ contributions, while the contributions from the $tS$, $tV$, and $tPS$ channels are negligible.
In other words, the total \textit{ph} strengths are determined by nothing but the delicate balance between the $\sigma$ and $\omega$ mesons (see Eq.~(\ref{Eq2})), which is well established in nuclear covariant density functionals.
Therefore, the constraints introduced by the Fock terms of the RHF scheme into the \textit{ph} residual interactions are straight forward and robust.

%\section{Summary and Perspectives}

\textbf{Summary and perspectives} In summary, a new local RHF equivalent covariant density functional was proposed recently in Ref.~\cite{Liang2012b}, where the constraints introduced by the Fock terms of the RHF scheme have been taken into account.
In this way, the advantages of existing RH functionals can be maintained, while the problems in the isovector channel can be solved.
This opens new and interesting perspectives for the development of nuclear local covariant density functionals with proper isoscalar and isovector properties.
To follow this direction, the investigation on the relativistic local density approximation for the Coulomb Fock terms is in progress \cite{Gu}.

%%%%%%%%%%%%%%%%%%%%%%%%%%%%%%%%%%%%%%%%%%%%%%%%
%% BACKMATTER
%%%%%%%%%%%%%%%%%%%%%%%%%%%%%%%%%%%%%%%%%%%%%%%%

\begin{theacknowledgments}
This work is partially supported by the Major State 973 Program 2007CB815000,
National Natural Science Foundation of China under Grant Nos. 10975008, 11105005, 11105006, and 11175002,
the Research Fund for the Doctoral Program of Higher Education under Grant No. 20110001110087,
the Overseas Distinguished Professor Project from Ministry of Education No. MS2010BJDX001,
the DFG Cluster of Excellence "Origin and Structure of the Universe" (www.universe-cluster.de),
and the JSPS Postdoctoral Fellowship for Foreign Researchers.
\end{theacknowledgments}

\bibliographystyle{aipproc}   % if natbib is available
%\bibliographystyle{aipprocl} % if natbib is missing

%%%%%%%%%%%%%%%%%%%%%%%%%%%%%%%%%%%%%%%%%%%
%% You probably want to use your own bibtex database here
%%%%%%%%%%%%%%%%%%%%%%%%%%%%%%%%%%%%%%%%%%%
%\bibliography{ref}

\begin{thebibliography}{23}
\expandafter\ifx\csname natexlab\endcsname\relax\def\natexlab#1{#1}\fi
\providecommand{\enquote}[1]{``#1''}
\expandafter\ifx\csname url\endcsname\relax
  \def\url#1{\texttt{#1}}\fi
\expandafter\ifx\csname urlprefix\endcsname\relax\def\urlprefix{URL }\fi
\providecommand{\eprint}[2][]{\url{#2}}

\bibitem[Liang et~al.(2012{\natexlab{a}})]{Liang2012b}
H.~Z. Liang, P.~W. Zhao, P.~Ring, X.~Roca-Maza, and J.~Meng, \emph{Phys. Rev.
  C} \textbf{86}, 021302(R) (2012{\natexlab{a}}).

\bibitem[Kohn and Sham(1965)]{Kohn1965}
W.~Kohn, and L.~J. Sham, \emph{Phys. Rev.} \textbf{140}, A1133 (1965).

\bibitem[Bender et~al.(2003)]{Bender2003}
M.~Bender, P.-H. Heenen, and P.-G. Reinhard, \emph{Rev. Mod. Phys.}
  \textbf{75}, 121 (2003).

\bibitem[Ring(1996)]{Ring1996}
P.~Ring, \emph{Prog. Part. Nucl. Phys.} \textbf{37}, 193 (1996).

\bibitem[Vretenar et~al.(2005)]{Vretenar2005}
D.~Vretenar, A.~V. Afanasjev, G.~A. Lalazissis, and P.~Ring, \emph{Phys. Rep.}
  \textbf{409}, 101 (2005).

\bibitem[Meng et~al.(2006)]{Meng2006}
J.~Meng, H.~Toki, S.-G. Zhou, S.~Q. Zhang, W.~H. Long, and L.~S. Geng,
  \emph{Prog. Part. Nucl. Phys.} \textbf{57}, 470 (2006).

\bibitem[Paar et~al.(2007)]{Paar2007}
N.~Paar, D.~Vretenar, E.~Khan, and G.~Col\`o, \emph{Rep. Prog. Phys.}
  \textbf{70}, 691 (2007).

\bibitem[Nik\v{s}i\'{c} et~al.(2011)]{Niksic2011}
T.~Nik\v{s}i\'{c}, D.~Vretenar, and P.~Ring, \emph{Prog. Part. Nucl. Phys.}
  \textbf{66}, 519 (2011).

\bibitem[Long et~al.(2006)]{Long2006}
W.~H. Long, N.~Van~Giai, and J.~Meng, \emph{Phys. Lett. B} \textbf{640}, 150
  (2006).

\bibitem[Roca-Maza et~al.(2011)]{Roca-Maza2011}
X.~Roca-Maza, X.~Vi\~nas, M.~Centelles, P.~Ring, and P.~Schuck, \emph{Phys.
  Rev. C} \textbf{84}, 054309 (2011).

\bibitem[Long et~al.(2007)]{Long2007}
W.~H. Long, H.~Sagawa, N.~Van~Giai, and J.~Meng, \emph{Phys. Rev. C}
  \textbf{76}, 034314 (2007).

\bibitem[Long et~al.(2008)]{Long2008}
W.~H. Long, H.~Sagawa, J.~Meng, and N.~Van~Giai, \emph{Europhys. Lett.}
  \textbf{82}, 12001 (2008).

\bibitem[Tarpanov et~al.(2008)]{Tarpanov2008}
D.~Tarpanov, H.~Z. Liang, N.~Van~Giai, and C.~Stoyanov, \emph{Phys. Rev. C}
  \textbf{77}, 054316 (2008).

\bibitem[Long et~al.(2009)]{Long2009}
W.~H. Long, T.~Nakatsukasa, H.~Sagawa, J.~Meng, H.~Nakada, and Y.~Zhang,
  \emph{Phys. Lett. B} \textbf{680}, 428 (2009).

\bibitem[Moreno-Torres et~al.(2010)]{Moreno-Torres2010}
M.~Moreno-Torres, M.~Grasso, H.~Z. Liang, V.~De~Donno, M.~Anguiano, and
  N.~Van~Giai, \emph{Phys. Rev. C} \textbf{81}, 064327 (2010).

\bibitem[Liang et~al.(2008)]{Liang2008}
H.~Z. Liang, N.~Van~Giai, and J.~Meng, \emph{Phys. Rev. Lett.} \textbf{101},
  122502 (2008).

\bibitem[Liang et~al.(2009)]{Liang2009}
H.~Z. Liang, N.~Van~Giai, and J.~Meng, \emph{Phys. Rev. C} \textbf{79}, 064316
  (2009).

\bibitem[Liang et~al.(2012{\natexlab{b}})]{Liang2012}
H.~Z. Liang, P.~W. Zhao, and J.~Meng, \emph{Phys. Rev. C} \textbf{85}, 064302
  (2012{\natexlab{b}}).

\bibitem[Ebran et~al.(2011)]{Ebran2011}
J.-P. Ebran, E.~Khan, D.~Pe\~na Arteaga, and D.~Vretenar, \emph{Phys. Rev. C}
  \textbf{83}, 064323 (2011).

\bibitem[Okun(1982)]{Okun1982}
L.~B. Okun, \emph{Leptons and Quarks}, North-Holland, Amsterdam, 1982.

\bibitem[Sulaksono et~al.(2003)]{Sulaksono2003}
A.~Sulaksono, T.~B\"{u}venich, J.~A. Maruhn, P.~G. Reinhard, and W.~Greiner,
  \emph{Ann. Phys. (NY)} \textbf{306}, 36 (2003).

\bibitem[Wakasa et~al.(1997)]{Wakasa1997}
T.~Wakasa, et~al., \emph{Phys. Rev. C} \textbf{55}, 2909 (1997).

\bibitem[Gu et~al.(2012)]{Gu}
H.-Q. Gu, et~al., in preparation.

\end{thebibliography}

\end{document}